# Giant Raman Intensity Modulation in Pristine Carbon Nanotubes

Adam W. Bushmaker[1], Vikram V. Deshpande[2], Scott Hsieh[2],
Marc W. Bockrath[2], Stephen B. Cronin[1]

[1]*University of Southern California, Department of Electrical Engineering.*
*Los Angeles, CA 90089*

[2]*California Institute of Technology, Department of Applied Physics*
*Pasadena, CA 91125*

**Abstract**

Large variations of up to two orders of magnitude are observed in the Raman intensity of pristine, suspended quasi-metallic single-walled carbon nanotubes in response to applied gate potentials. No change in the resonance condition is observed, and all Raman bands exhibit the same changes in intensity, regardless of phonon energy or laser excitation energy. The effect is not observed in semiconducting nanotubes. The electronic energy gaps correlate with the drop in the Raman intensity, and the recently observed Mott insulating behavior is suggested as a possible explanation for this effect.

Single-walled carbon nanotubes (SWNTs) provide an excellent system for studying interesting one-dimensional physics, including exceptionally strong electron-phonon coupling [1-3], ballistic electron transport [4], and strongly correlated electrons [5-8]. Micro-Raman spectroscopy has proven to be a sensitive technique for observing these unique effects [9, 10]. Despite the great interest and large volume of literature on SWNTs, new phenomena such as those mentioned above are still being discovered with the use of clean, nearly defect free, suspended SWNTs. Understanding these effects in pristine systems is crucial for the future development of nano-devices based on metallic SWNTs.

It is well known that the Raman intensity of SWNTs is significantly enhanced when one of the photons involved is resonant with an excitonic transition [11-15]. There have been several reports on slight changes in the Raman spectral intensity from SWNTs in response to gate voltages, which were attributed to shifting of the resonance condition [16, 17], as well as reports on larger intensity changes in SWNTs under extreme electrolytic doping, due to transition bleaching [18, 19] or otherwise anomalous behavior in complex nanotube mats [20]. Raman studies of electrostatically doped graphene have also been undertaken [21, 22], showing moderate decreases in the *2D* band Raman intensity with doping [22].

In this study, the Raman spectra of individual, suspended, pristine quasi-metallic (small bandgap or "qm") SWNTs are found to exhibit an increase in intensity by up to two orders of magnitude with an applied electrostatic gate voltage, while for semiconducting nanotubes the intensity remains constant. As such, the effect may be used as a means to identify pristine metallic nanotubes. The effect is so strong that it renders some qm-SWNTs invisible to Raman spectroscopy, and occurs over small voltage ranges, suggesting possible device applications in the future. In contrast to the previous work [16-20], we observe an *increase* in intensity with doping, as opposed to a decrease. Furthermore, this increase occurs with relatively small gate voltages, in contrast with other studies that used several volts of electrolytic doping or several tens of volts with electrostatic doping. Changes in the resonance condition are ruled out based on the invariance of this effect with respect to phonon energy, laser energy, and Stoke/anti-Stokes intensity ratio. By performing optical and electrical measurements simultaneously, the electrically measured energy gaps ($E_{gap}$) are compared to the FWHM drops in Raman intensity. Based on these results, the recently observed Mott insulating behavior [23] in qm-SWNTs is suggested as a possible mechanism for the observed Raman intensity modulation.

Recently, there has been a large focus on the Raman $G_-$ band's response to applied gate voltages [10, 20, 24-27]. In these studies, the $G_-$ band frequency and linewidth change drastically due to the influence of the $\Gamma$-point Kohn anomaly in the LO phonon band [28]. These effects were also observed in our devices, and were reported previously [27]. The observed intensity modulation, reported here, affects all Raman modes universally, not just those associated with the Kohn anomaly. Furthermore, in bands not affected by the Kohn anomaly, no noteworthy shifts or changes in linewidth are observed.

Samples are fabricated using chemical vapor deposition on Pt electrodes with predefined catalyst beds, as reported previously [9, 29]. The resulting devices are single-walled nanotubes suspended across trenches 300nm deep and 2-5 µm wide (see Figure 1a). The samples in this study were grown using ethanol or methane as the carbon feedstock [30]. No additional processing was performed after the nanotube growth, except for an oxygen bake to rid the devices of amorphous carbon. The devices are screened by examination of the Raman and electrical characteristics. All nanotubes in this study exhibited a single spatially isolated Raman signal, and a high bias saturation current of ~10/L (µA), where L is the length in microns [2]. The low temperature transport data from most devices exhibited coulomb blockade diamonds [5], and all of the devices exhibited little or no $D$ band Raman intensity. These observations indicate that all nanotubes in this study are highly defect-free, individual qm-SWNT devices. Raman spectra were collected from the center of each nanotube in the middle of the trench with a Renishaw InVia spectrometer (resolution ~1cm$^{-1}$) using 532nm, 633nm, or 785nm lasers focused to a diffraction limited spot.

*G* band Raman spectra taken with a 785nm laser from an individual, suspended, qm-SWNT are plotted in Figure 1b at several gate voltages. As with all qm-SWNTs measured in this study, the intensity of the Raman signal increases dramatically with increasing $|V_g|$, varying by up to almost two orders of magnitude (>18.8 dB) in this case. Here, the $G_+$ and $G_-$ bands exhibited an identical intensity change. The *G* band lineshape in Figure 1b is typical of quasi-metallic nanotubes, exhibiting a broad, downshifted $G_-$ band, with a sharp $G_+$ band. Note the near-absence of the defect-related *D* band. A radial breathing mode (RBM) for this nanotube was observed at 173.6±0.5 cm$^{-1}$, indicating that the diameter of this SWNT is 1.31 nm [31].

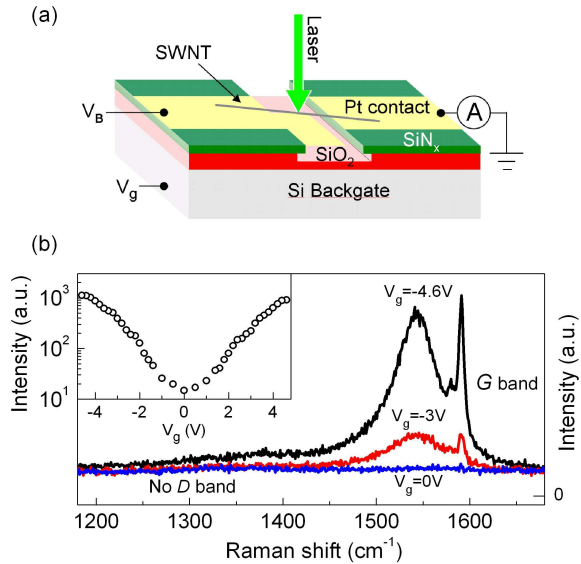

**Figure 1:** (a) Device geometry and (b) *G* band Raman spectra at various gate voltages, with inset showing the $G_-$ band intensity as a function of gate voltage (sample 8B).

Figure 2 shows the Raman data for another nanotube device, including the RBM, $G_+$ band, and *G'* band Raman intensities plotted as a function of $V_g$ and the Fermi energy ($E_F$), fit from the electrical data (discussed below). The normalized Raman intensity profiles show nearly identical gate voltage dependences, indicating that this effect affects all of the Raman modes universally, regardless of phonon energy. The $G_-$ band also exhibited the same dependence, but is not shown in the plot to maintain clarity.

The RBM, observed at 153±0.5 cm$^{-1}$ using both 633nm and 785nm lasers, shows similar intensity profiles (Figure 2b), with a Raman signal attenuation of 8.5 dB at $V_g = 0$. Throughout the measurement, the intensity of the background Si Raman band at 520 cm$^{-1}$ remained constant. Also shown in the figure is the temperature normalized (300K) RBM anti-Stokes/Stokes (AS/S) intensity ratio, which is known to be very sensitive to any changes in the resonance condition [32].

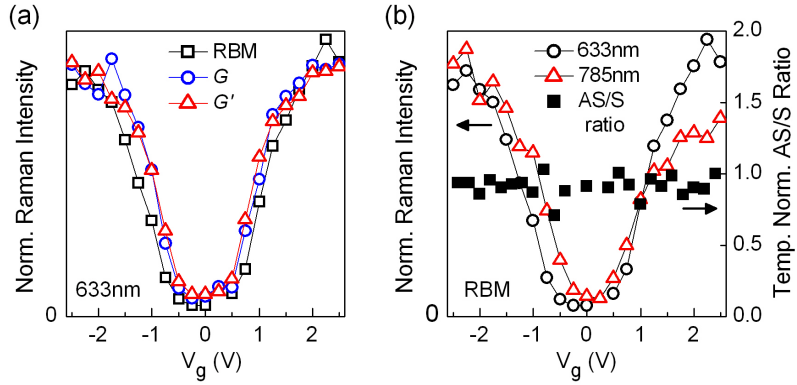

**Figure 2:** (a) Normalized Raman intensities of the RBM, $G_+$ (TO), and $G'$ bands taken with a 633nm laser. (b) RBM intensity taken with 633 and 785nm lasers, together with the RBM AS/S intensity ratio (normalized for T=300K) taken with the 633nm laser, plotted as a function of gate voltage. (sample 18B)

Normally, any changes in the Raman intensity of carbon nanotubes would be due to a change in the resonance condition. However, we find this not to be the case for these pristine, suspended qm-SWNTs. The resonant Raman intensity of the Stokes process is given by

$$I^\mu(E_L) = C \left| \int \frac{M^{op} M^{ep} M^{op} dk}{[E_L - E_\mu(k) - i\gamma][E_L - E_\mu(k) - E_{ph} - i\gamma]} \right|^2, \quad (1)$$

where $C$ is a constant, $E_L$ is the laser energy, $E_\mu$ is the excitonic transition energy between the $\mu^{th}$ valence and conduction subbands, $E_{ph}$ is the phonon energy, $\gamma$ is the resonance broadening energy, $M^{op}$ is the optical matrix element for the exciton-photon interaction, and $M^{ep}$ is the electron-phonon coupling matrix element [32]. A large change in the Raman intensity can arise from a change in three quantities: 1.) the resonance condition $|E_L - E_\mu|$ (or $|E_L - E_\mu - E_{ph}|$ ), 2.) $M^{ep}$, or 3.) $M^{op}$.

We can rule out transition bleaching immediately, as the changes in Fermi energy ($\leq$200meV) are drastically smaller than the energy of the excitonic transitions (1.5-2.3eV). It is tempting to attribute the change in Raman intensity to a strain-induced change in the resonance condition (case 1) caused by the electrostatic gate force [33]. However, this is not the case, since the RBM has a narrow resonance window, and, therefore, small changes in the resonance condition $(E_\mu - E_L)$ result in large changes in the RBM AS/S intensity ratio [32], which are not observed (Figure 2b). Also, the broad G band resonance window would require an unreasonably large change in $E_\mu$ in order for such a drastic modulation of the Raman intensity to take place. Therefore, we would expect the Raman signal for different phonon modes and different laser energies to respond differently to a change in resonance condition, which is not observed (Figures 2a and 2b). Finally, it is statistically unlikely that we would observe a shift *onto* resonance with increasing $|V_g|$ for all 8 nanotubes showing this effect. One would expect there to be some nanotubes showing a shift *off* of resonance with increasing $|V_g|$. The unanimous evidence in this respect suggests that a different mechanism is responsible for the observed behavior. We can rule out gate voltage-induced bending as a cause for the observed intensity modulation, as most suspended nanotubes have slack (and thus bending) as fabricated [34], and since no strain is observed one would expect no bending. Furthermore, the Raman intensity is predicted to decrease with bending [35], the opposite of the observed behavior.

Ruling out the denominator of Equation (1) as a possible explanation for the observed intensity modulation, we consider the electron-phonon coupling strength, $M^{ep}$ (case 2), which is known to be quite different for the various Raman active modes [36]. Therefore, a variation of this quantity is expected to result in different intensity modulation profiles for the RBM, G and G' bands, which is not observed (Figure 2a). This is especially true with the $G_+$ and $G_-$ bands, which have orthogonal TO and LO polarizations in qm-SWNTs, respectively [28]. The electron-phonon coupling of the LO phonon band is heavily influenced by the Kohn anomaly [10, 20, 24-28], and is drastically different from that of the TO phonon band. Despite this large difference in electron-phonon coupling strengths, the data show no significant difference in the intensity behavior of the TO and LO ($G_+/G_-$) phonon modes. This leaves a change in the optical matrix element $M^{op}$ (case 3) as the only plausible cause of the observed intensity modulation. This intensity modulation appears to be an attenuation at small $|V_g|$, rather than an amplification (or enhancement) at high $|V_g|$, because the Raman intensity saturates at high $|V_g|$ to a constant value comparable to that of the semiconducting SWNTs.

A Raman intensity map of the *G* band of a third nanotube is plotted in Figure 3a, together with the electrically measured conductance. In this intensity map, the *G* band peaks around 1580cm$^{-1}$ vanish near $E_F = 0$. This corresponds to the drop in the conductance observed in the electrical data. The conductance is modeled using the Boltzmann-Landauer (BL) transport equation [27, 37], and the Fermi energy is calculated numerically as a function of gate voltage using a geometric gate capacitance *C*, the Fermi function, and a hyperbolic density of states model [38], according to the equation $E_F + \frac{Q(E_F)}{C} = eV_g$, where *Q* is the charge induced on the nanotube. This accounts for the quantum capacitance [39] and the effect of the bandgap, which create the non-linear $V_g$-$E_F$ relationship shown in Figure 4b. Fitting the data in Figure 3a with this model yielded C ~10pF/m and $E_{gap} = 120$meV. The small offset of the conductance and Raman intensity minima near $V_g = 0$ arises from the gas doping effect at the electrodes [40]. This nanotube exhibits Raman attenuation below voltages of 2V and, as with the others, saturation of the Raman intensity at large gate voltages. Interestingly, we do not observe this same effect in semiconducting SWNTs (Figure 3b), which have bandgaps on the order of 1eV. Therefore, this Raman intensity modulation is not simply due to a change in the free carrier density.

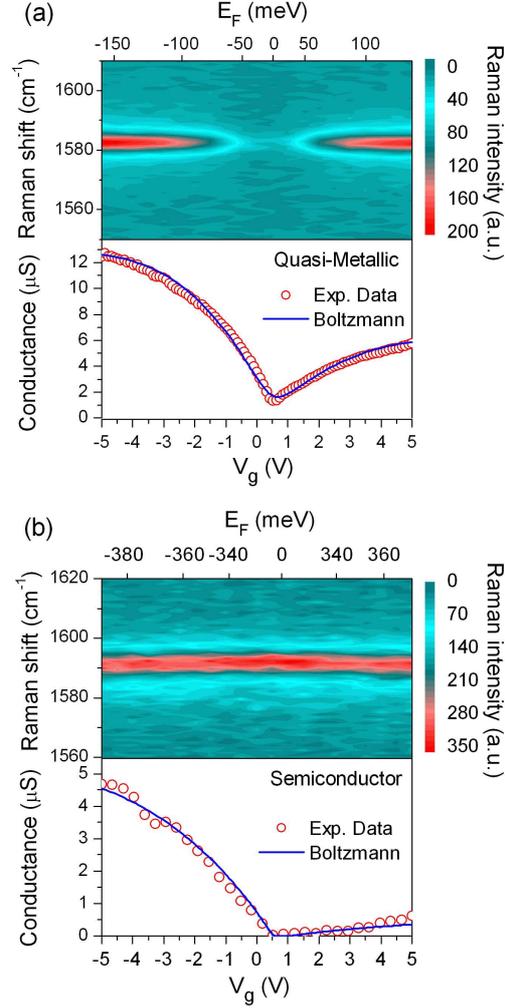

**Figure 3:** The *G* band Raman spectra and conductance for (a) quasi-metallic and (b) semiconducting suspended SWNTs, plotted versus gate voltage and Fermi energy. The electrical data and Fermi energies are fit using the Boltzmann-Landauer transport equation. (sample 22B2 (m) and 16B (sc))

Out of 9 qm-SWNTs investigated, 8 showed this giant intensity modulation effect. The Raman intensity of the remaining qm-SWNT was constant. A total of 4 semiconducting SWNTs were also investigated using this technique, none of which showed substantial Raman intensity changes with gate voltage. We, therefore, conclude that the intensity modulation effect is specific to qm-SWNTs. The data for the 8 qm-SWNTs showing this effect are summarized in Table 1 below. The diameter is given for nanotubes that exhibited a RBM in their Raman spectra, calculated using the relation $d_t = 227/\omega_{RBM}$ [31].

The $G_+/G_-$ Raman integrated intensity ratios are also listed for each nanotube in the table, and give an indication of the chiral angle ($G_+/G_- = 0$ → zigzag, $G_+/G_- = \infty$ → armchair) [25, 41]. Also given are the maximum observed Raman attenuation (in dB) and the Fermi energy change corresponding to the FWHM attenuation of the Raman intensity ($\Delta_{Raman}$), found using the $V_g$-$E_F$ relationship, which is outlined in Figure 4b. Finally, the energy gaps ($E_{gap}$) obtained by fitting the BL model to the measured conductance are also given for each qm-SWNT. Figure 4a shows $\Delta_{Raman}$ plotted versus $E_{gap}$. The correlation between $\Delta_{Raman}$ and $E_{gap}$ suggests that the observed Raman intensity attenuation is caused by the same effect that causes the electronic energy gaps in qm-SWNTs.

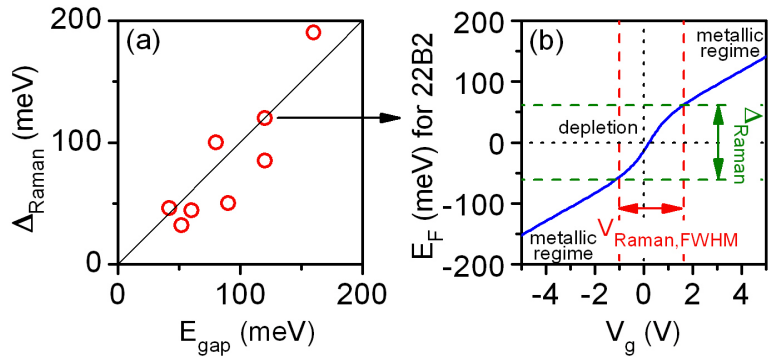

| Sample | Diameter (nm) | G+/G- Int. Rat. | Att. (dB) | $\Delta$Raman (meV) | Egap (meV) |
|---|---|---|---|---|---|
| 22B1 | - | 0.04 | 12.8 | 50 | 90 |
| 13B | - | 0.12 | 0.8 | 85 | 120 |
| 8B | 1.31 | 0.08 | >18.8 | 190 | 160 |
| 18B | 1.48 | 0.40 | 8.5 | 44 | 60 |
| 22B2 | - | → ∞ | 11.8 | 120 | 120 |
| 19B | - | → ∞ | 11.5 | 32 | 52 |
| 5A | 1.97 | → 0 | 13.8 | 46 | 42 |
| 5B | - | → 0 | 7.3 | 100 | 80 |

**Table 1 and Figure 4:** Data summary of qm-SWNTs showing intensity modulation. Listed values include nanotube diameter, $G_+/G_-$ integrated Raman intensity ratio, maximum Raman attenuation, attenuation energy gap ($\Delta_{Raman}$), and the electronic energy gap ($E_{gap}$). In Figure 4a, $\Delta_{Raman}$ is plotted versus $E_{gap}$, determined by fitting the Boltzmann-Landauer transport equation to the experimental transport data. In Figure 4b, $E_F$ is plotted versus $V_g$ for nanotube 22B2, illustrating the method for determining $\Delta_{Raman}$.

The secondary bandgap in qm-SWNTs (those SWNTs with chiral indices such that n-m is an integer multiple of 3) has long been though to arise from the curvature of the nanotube, which causes mixing of the π and σ orbitals [41, 42]. A Peierls gap transition, one hallmark feature of most one-dimensional metals, was initially considered as a possible cause for the electrical bandgap. However, density functional theory (DFT) investigations have found the Peierls gap to be unstable above T~$10^{-8}$ K [28] in all but ultrasmall radius carbon nanotubes [43, 44]. Recently, experimental evidence [23] has confirmed theoretical predictions [45, 46] that, in nearly defect-free qm-SWNTs, a Mott insulator transition is primarily responsible for creating $E_{gap}$. In the Mott insulating state, strongly correlated electrons localize to their parent atoms, forming gaps of 10-100 meV, even in armchair SWNTs. Raman intensity attenuation has been previously reported for Mott insulator transitions in other materials systems [47]. We believe that this same effect is causing the Raman attenuation in these nearly defect free nanotubes. The fit values for $E_{gap}$ in Table 1 lie in the range predicted for Mott insulating gaps, and correlate well with the energy gaps over which the Raman attenuation is observed (Figure 4a), corroborating the doping mediated Mott insulator state.

The Mott insulator transition explains why all the Raman bands are affected equally under applied gate potentials. In this phase transition, the electrons in the *2p*-orbital of the carbon atom localize to their parent atom through Coulomb repulsion, causing all the electrons in the π-band to be affected, including those involved in excitonic transitions. The details of this interaction are left to future theoretical work. The Mott insulator transition also explains the specific occurrence in quasi-metallic nanotubes, as opposed to semiconductors, since the Mott insulator occurs only in quasi-metallic nanotubes. In semiconductor nanotubes, the electronic bandgap originates from confinement effects. Absorption studies (optical [48] and X-ray [49]) in other materials systems have also shown dramatic changes as a result of the Mott transition. Finally, the gate voltage-induced Mott insulator transition has already been exploited in cuprate Mott transition field effect transistors (MTFETs) [50]. It is likely that this modulation has not been observed until now because most gate voltage experiments with qm-SWNTs are performed on nanotube-on-substrate devices, rather than pristine, suspended devices. The Mott insulating state requires the presence of a well-defined charge neutrality point [23], which may not occur in samples with defects, substrate contact, or post-processing residue.

In conclusion, we observe a large attenuation of the Raman signal from individual pristine, suspended quasi-metallic SWNTs by up to two orders of magnitude near zero electrostatic gating, while semiconducting nanotubes do not exhibit the effect. The attenuation is so strong as to render some qm-SWNTs undetectable by Raman spectroscopy in the absence of an applied gate voltage. Changes in the resonance condition and transition bleaching are ruled out on the basis of the constant anti-Stokes/Stokes intensity ratio and the universal character of the effect with respect to different phonon modes and laser energies. The changes are attributed to attenuation of the optical matrix element and the recently observed Mott insulator transition in qm-SWNTs is suggested as a possible mechanism. The Raman attenuation energy gaps for 8 nanotubes are compared to the electronic energy gaps, estimated from fits to the Boltzmann-Landauer transport model, and show correlation consistent with the Mott insulator picture.

This research was supported in part by DOE Award No. DE-FG02-07ER46376 and the National Science Foundation Graduate Research Fellowship Program.